\documentclass[10pt,conference]{IEEEtran}

\usepackage{cite}
\usepackage{color}
\usepackage{graphicx}
\usepackage{xcolor}
\usepackage{listings}
\usepackage{tabularx}
\usepackage{booktabs}
\usepackage{array}
\usepackage{colortbl}
\usepackage{longtable}
\usepackage{seqsplit}
\usepackage{hyperref}

\usepackage{xurl}
\usepackage{float}
\usepackage{multirow}
\usepackage{listings}
\usepackage[most]{tcolorbox}

\newtcolorbox{answerbox}{colback=gray!8, colframe=black, boxrule=0.5pt, arc=1pt, left=6pt, right=6pt, top=4pt, bottom=4pt}

\newcolumntype{L}{>{\raggedright\arraybackslash}X}
\newcolumntype{P}[1]{>{\raggedright\arraybackslash}p{#1}}

\newcommand{\TODO}[1]{\textcolor{red}{#1}\GenericWarning{}{LaTeX Warning: TODO: #1}}\newcommand\todo\TODO

\definecolor{listingbg}{gray}{0.94}

\definecolor{cellgreen}{HTML}{A3D9AD}
\definecolor{celllightgreen}{HTML}{E4F5E7}
\newcommand{\cellfrac}[2]{%
  \ifnum#2>0
    \ifnum#1=#2 \cellcolor{cellgreen}%
    \else\ifnum#1>0 \cellcolor{celllightgreen}\fi
    \fi
  \fi
  #1/#2}
\makeatletter
\def\lst@makecaption#1#2{\vbox{\@makecaption{#1}{#2}}\vskip\belowcaptionskip}
\makeatother

\lstdefinestyle{promptstyle}{
  backgroundcolor=\color{listingbg},
  basicstyle=\small\ttfamily,
  breaklines=true,
  breakindent=0pt,
  frame=single,
  framesep=4pt,
  rulecolor=\color{gray!40},
  xleftmargin=4pt,
  xrightmargin=4pt,
  columns=fullflexible,
}

\lstdefinestyle{inline}{
   basicstyle=\small\ttfamily,
   keywordstyle=\color{black},
   backgroundcolor=\color{yellow},
   breaklines=true,
}

\lstdefinestyle{mystyle}{
    basicstyle=\footnotesize\ttfamily, %
    numberstyle=\tiny, %
    numbersep=5pt,%
    tabsize=2,  %
    extendedchars=true,
    breaklines=true,  %
    keywordstyle=\color{red},
    frame=lines,
    columns=fullflexible,
    stringstyle=\color{white}\ttfamily, %
    showspaces=false,
    showtabs=false,
    xleftmargin=5pt,
    framexleftmargin=5pt,
    framexrightmargin=5pt,
    framexbottommargin=5pt,
    showstringspaces=false,
    moredelim=**[is][\color{red}]{~}{~}, 
    emph={@kwsites/file-exists@ˆ1.1.1}, 
    emphstyle=[1]\color{red} 
}

\definecolor{yulred}{HTML}{C0392B}
\definecolor{yulgreen}{HTML}{2F9E44}
\definecolor{yulgrey}{HTML}{57606A}
\lstdefinestyle{yulstyle}{
  backgroundcolor=\color{listingbg},
  basicstyle=\scriptsize\ttfamily,
  breaklines=true,
  breakatwhitespace=true,
  breakindent=1em,
  frame=single,
  framesep=4pt,
  rulecolor=\color{gray!40},
  xleftmargin=4pt,
  xrightmargin=4pt,
  columns=fullflexible,
  keepspaces=true,
  moredelim=[il][\color{black}\bfseries]{|},
  moredelim=[is][\bfseries]{[*}{*]},
  moredelim=[is][\color{yulred}\bfseries]{[!}{!]},
  moredelim=[is][\color{yulred}\bfseries]{[-}{-]},
  moredelim=[is][\color{yulgreen}\bfseries]{[+}{+]},
  literate={->}{{\textcolor{black}{$\rightarrow$}}}2,
}

\makeatletter
\def\ext@lstlisting{lol}
\newenvironment{listingfloat}[1][tbp]{\@float{lstlisting}[#1]\def\fnum@lstlisting{\lstlistingname~\thelstlisting}}{\end@float}
\makeatother
\newcommand{\sublisting}[1]{\par\vspace{4pt}{\footnotesize #1}\par\vspace{1pt}}

\usepackage{xspace}

\newcommand{\tool}{\textsc{yul}\xspace}

\newif\ifanonymous
\anonymousfalse
\ifanonymous
\else
  \newcommand{\yulurl}{\url{https://github.com/chains-project/yul}}
  \newcommand{\benchmarkurl}{\url{https://github.com/chains-project/yul-benchmark}}
\fi

\begin{document}
\pagestyle{plain}

\title{\tool: Bringing Software Supply Chain Security to Coding Agents}

\ifanonymous
\author{\IEEEauthorblockN{Anonymous Author(s)}}
\else
\author{
\IEEEauthorblockN{Aman Sharma, Frank Reyes, Martin Monperrus}
\IEEEauthorblockA{KTH Royal Institute of Technology, Stockholm, Sweden}
}
\fi

\maketitle
\thispagestyle{plain}

\begin{abstract}
Coding agents choose and declare third-party dependencies on their own.
They often declare a stale version that they recall from their training data.
A stale dependency exposes the project to known vulnerabilities that the latest release already fixes.
We present \tool, a design for coding agent harnesses that ensures that no stale dependencies are ever used.
It does so by checking every dependency declaration against the package registry.
If a dependency is stale or does not exist, \tool blocks the usage and tells the coding agent which version to use, so the coding agent retries with the fix.
We evaluate \tool on 60 tasks across six ecosystems with Claude Code running Claude Opus 5.5 and OpenCode running DeepSeek V4.1 Flash.
Without \tool, Opus declares a stale version in 29\% of runs and DeepSeek in 39\%, and both declare a stale version in every GitHub Actions run.
With \tool, the coding agents end on the latest release in 93\% of stale Opus runs and in 96\% of stale DeepSeek runs.
\tool thus brings the best practices of software supply chain security to coding agents.
\end{abstract}


\section{Introduction}

Coding agents solve programming tasks on their own.
A developer describes a task in natural language, and the coding agent plans a solution, edits files, runs commands, and tests the result until the task is complete.
Along the way, coding agents do an essential software engineering task: they choose third-party dependencies to reuse.

When a task needs a dependency, the coding agent declares it in a manifest, such as \texttt{package.json} or \texttt{pom.xml}, and a version is taken.
This dependency version can be vulnerable or even malicious, and it then puts the whole project at risk~\cite{ohm_backstabber_2020}.
In this paper, we address the problem that coding agents have no principled way of using dependencies.

First and foremost, coding agents may use stale dependencies.
A dependency is stale when its declaration does not resolve to the latest release of its package.
This happens for instance when the coding agent recalls a version from the training data, which ends at a knowledge cutoff.
Stale dependencies carry a security cost, since systems with outdated dependencies are four times as likely to have security issues as systems that are up to date~\cite{cox_freshness_2015}.
The OpenSSF guide for AI code assistants recommends a prompt that tells the coding agent to ``prefer the latest stable release''~\cite{openssf_guide_2025}.
This is weak:  prompting alone cannot enforce that only latest versions are used and agents may still use a stale version.

To solve this problem, we present \tool, a coding agent design pattern that guarantees that only latest dependency versions are used. 
\tool looks up the latest release in the package registry for each dependency that a coding agent wants to use.
If the coding agent declares a stale version, \tool blocks the write and reports the latest release, so the coding agent adapts its behavior.

To evaluate \tool, we build a benchmark of 60 dependency-aware coding tasks across six ecosystems, namely Maven, GitHub Actions, PyPI, npm, Go modules, and Cargo.
A task asks for a feature but never mentions a dependency, so that the coding agent can decide on its own what to reuse and which version to pick.
We run each task three times through two coding agents, Claude Code with Claude Opus 5.5 and OpenCode with DeepSeek V4.1 Flash, and measure how often they declare a stale version on their own (RQ1).
Finally, we measure how effectively \tool mitigates the stale versions (RQ2).

Our results show that \tool is effective in mitigating the stale dependency problem.
Without \tool, Opus declares a stale version in 42 of 146 valid runs (29\%) and DeepSeek in 54 of 139 (39\%) cases.
With \tool, the coding agents end on the latest release in 39 of 42 stale Opus runs (93\%) and in 52 of 54 stale DeepSeek runs (96\%). The remaining cases where \tool does not prevent the usage of stale dependencies ties to corner cases of environment management and dependency resolution.

To summarize, this paper makes the following contributions.
\begin{itemize}
  \item An empirical study of how often coding agents declare and use stale dependency versions across six ecosystems.
  \item \tool, a publicly available system for coding agents that blocks stale dependency declarations, available at \yulurl.
  \item An evaluation of \tool on 60 tasks and 720 runs, which shows that \tool mitigates 93\% and 96\% of the stale dependency cases.
  \item A benchmark of 60 dependency tasks and the transcripts of all runs, available at \benchmarkurl.
\end{itemize}

Sections~\ref{sec:background} to~\ref{sec:conclusion} present, in order, the background, the design of \tool, the methodology, the results, a discussion, threats to validity, related work, and the conclusion.

\section{Background}
\label{sec:background}

\subsection{Coding Agents}
\label{bkg:agents}
A \emph{coding agent} combines an LLM with a \emph{harness}, which runs the LLM in a loop with specific prompting and format expectations.
Examples of coding agents include Claude Code or OpenCode. 
When a coding agent is given a coding task, it leverages the LLM to plan, write, and test code until the task is complete.

A harness lets users run their own commands, called \emph{hooks}, at fixed points of this loop~\cite{anthropic_hooks_2026}.
Hooks are not prompting, they give deterministic control, by being run in the loop instead of relying on the judgment of the LLM to run them~\cite{anthropic_hooksguide_2026}.
For example, developers can use hooks to format code after each edit by the LLM.
A hook that runs before a tool call can allow the call or block it.
When the hook blocks a call, the LLM receives the message of the hook instead of the result of the call, and the coding agent can retry with a corrected call.
There are different ways of implementing hooks, Claude Code supports external programs as hooks, while OpenCode offers a plugin system~\cite{opencode_plugins_2026}.

\subsection{Software Supply Chain Security}
\label{bkg:ssc}
Virtually all software projects reuse third-party packages, which form the \emph{software supply chain} of the project~\cite{soto-valero_multibillion_2022}.
A project declares these packages as dependencies in a \emph{manifest}, such as \texttt{package.json} for npm or \texttt{pom.xml} for Maven, and the package manager downloads them from the \emph{registry} of the ecosystem.

Each dependency is code that the developers of the project do not write.
If the dependency is vulnerable or malicious, it can put the whole project at risk, subject to a software supply chain attack.
Software supply chain security is about practices and tooling to protect against such dependency risk~\cite{pashchenko_qualitative_2020}.

Dependency updates are a particular risk surface.
An update brings a project the latest features and fixes of its packages, including fixes for known vulnerabilities.
A missed update keeps these vulnerabilities in the project, even though a fix already exists.
We call a dependency \emph{stale} when its declaration does not fetch its latest release.
The \emph{latest release} of a package is the most recent stable release that its registry reports.
For example, \texttt{requests==2.31.0} is stale if the latest release is \texttt{2.34.2}, while \texttt{requests>=2.31.0} is not stale, because it permits the installation of any latest release.
A stale dependency is a security risk, because it exposes the project to vulnerabilities that are known and exploited~\cite{cox_freshness_2015}.
For this reason, using automated update bots such as Dependabot and Renovate is considered a best practice~\cite{zahan_openssf_2023}.

Yet update bots leave a huge gap.
A bot runs on the repository on a schedule, so it only sees a stale dependency after a developer pushes it.
Between the moment a developer installs a stale dependency on their machine and the moment the update of the bot is merged, even automatically, the project builds and runs on the stale version.
Cooldowns futher widen this, because they delay each update of the bot by days~\cite{tanaka_exploratory_2026}.
Coding agent make the problem even more acute, because they write the stale version into the manifest and build and run code on top of it right away.
This is exactly the gap that we address in this paper: the window in which stale dependencies are downloaded and executed.

\section{\tool Design}
\label{sec:design}

\subsection{Overview}

We propose a design in order to guarantee that coding agents do not use stale dependencies.
Our prototype, \tool, implements this design as an extension to an existing coding agent harness.
The core idea of \tool is to intercept stale dependencies as early as possible, as soon as they come out of the language model.

\begin{figure*}[!t]
  \centering
  \includegraphics[width=\textwidth]{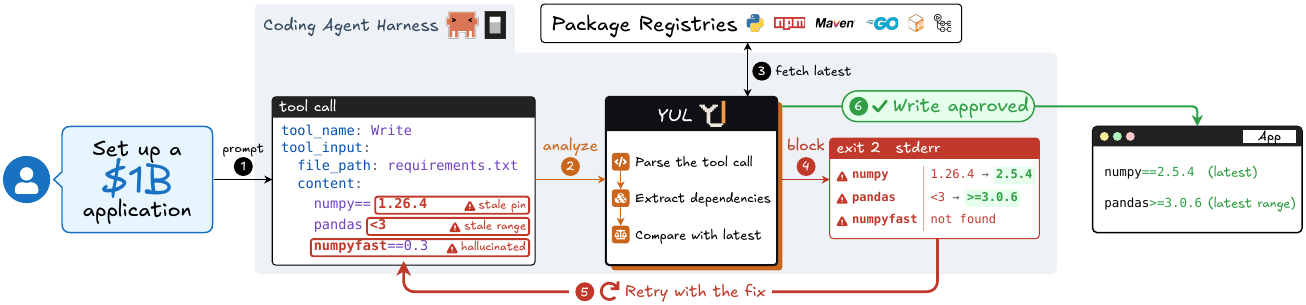}
  \caption{Overview of \tool, a tool that guarantees that coding agents never use stale dependencies.}
  \label{fig:overview}
\end{figure*}

To guarantee the absence of stale dependencies, \tool works in six steps, as Figure~\ref{fig:overview} shows.
(1) A developer prompts a coding agent to perform a programming task.
The coding agent identifies the need for a library, and thus updates a manifest to import the library.
(2) \tool intercepts the manifest write before it reaches disk and analyzes it.
(3) \tool looks up the latest release of each dependency in the corresponding package registry.
(4) If a dependency is stale or does not exist, \tool blocks the write and tells the coding agent how to fix it.
(5) This forces the coding agent to retry the write with the fix.
(6) Once every dependency is up to date, \tool approves the write and lets it reach disk.
As a result, \tool guarantees that no stale dependency is ever written, downloaded and executed.\footnote{Note that simply asking the agent to use the latest version via prompting sometimes works but never gives you guarantees.}

\subsection{Input}
\tool receives the manifest that the coding agent is about to write.
For example, in Figure~\ref{fig:overview}, the coding agent wants to write a Python \texttt{requirements.txt} that pins \texttt{numpy} 1.26.4, restricts \texttt{pandas} to releases below 3, and declares \texttt{numpyfast}.
Every dependency in this manifest is a problem that \tool mitigates.
The pin on \texttt{numpy} 1.26.4 is stale, because the latest release is 2.5.4.
The range on \texttt{pandas} is stale, because it excludes the latest release 3.0.6.
The package \texttt{numpyfast} is an hallucination that does not exist on PyPI.
A hallucinated name is a supply chain risk, because an attacker can register it and publish malicious code under it~\cite{spracklen_package_2025}.

A coding agent declares a dependency in the manifest in one of two ways.
A \emph{file write} saves the new content of the manifest directly, as in Figure~\ref{fig:overview}.
A \emph{package manager command} records the dependency in the manifest, from a command emitted by the coding agent, without direct file write.
For example, \texttt{npm install react@19.3.1} installs React 19.3.1 and records it in \texttt{package.json}, even if a newer release exists.
\tool intercepts both cases. For a command, it checks the version that the command pins before the shell runs it.
A command without a version, such as \texttt{npm install react}, installs the latest release, so \tool lets it pass.

\subsection{Workflow}
\tool extracts every dependency from the manifest and compares it with the latest release in the package registries, such as PyPI or Maven Central.
\tool blocks a dependency in three cases.
(1) \tool blocks an exact pin on a release older than the latest release, such as \texttt{numpy==1.26.4} when the latest release is 2.5.4, and suggests the latest release as the fix.
(2) \tool blocks stale dependencies (\autoref{bkg:ssc}) and ranges that excludes the latest release, such as \texttt{pandas<3} when the latest release is 3.0.6, and suggests a range that starts at the latest release, such as \texttt{>=3.0.6,<4.0.0}.
(3) \tool blocks a dependency that does not exist in its registry, such as \texttt{numpyfast}, it is an hallucination. \tool has no release to suggest for hallucination, so it simply asks the coding agent to replace it with a real package.
\tool accepts every other dependency addition, which is either an exact pin on the latest release or a range that admits the latest release.
For example, \tool accepts \texttt{numpy==2.5.4} and \texttt{pandas>=2}.

\subsection{Output}
If \tool finds a stale or hallucinated dependency, it blocks the write and tells the coding agent how to fix each dependency.
Listing~\ref{lst:yul-output} shows this message for the write in Figure~\ref{fig:overview}.
For a hallucinated dependency, \tool asks the coding agent to use a real package instead, as Listing~\ref{lst:yul-output}a shows.
For a stale pin, \tool suggests the latest release, such as \texttt{numpy} 2.5.4 instead of 1.26.4, as Listing~\ref{lst:yul-output}b shows.
For a stale range, \tool suggests a range that admits the latest release, such as \texttt{>=3.0.6,<4.0.0} instead of \texttt{<3} for \texttt{pandas}.
Then, the coding agent reads the message and retries another write as part of the agentic loop.
\tool checks every retry again, so it also blocks a retry that ignores the fix.

In Figure~\ref{fig:overview}, \tool finds that \texttt{numpy} 1.26.4 is a stale pin, that \texttt{pandas<3} is a stale range, and that \texttt{numpyfast} does not exist on PyPI. In that case, \tool blocks all of them as long as the LLM has not fixed all three issues. 
When the LLM finally uses the latest version, \tool accepts the manifest to be written to disk.

\begin{listingfloat}[t]
\caption{Message from \tool to the coding agent for the write in Figure~\ref{fig:overview}.}
\label{lst:yul-output}
\sublisting{(a) \tool response for hallucinated dependency}
\begin{lstlisting}[style=yulstyle]
|replace these dependencies with real packages; they don't exist on their registry, so they're likely hallucinated:
  [!numpyfast!]  [-0.3-]
\end{lstlisting}
\sublisting{(b) \tool response for outdated dependencies}
\begin{lstlisting}[style=yulstyle]
|outdated dependencies, use these versions instead:
  [*numpy*]  [-1.26.4-] -> [+2.5.4+]
|these pinned ranges need attention:
  [*pandas*]  [-<3-] does not allow latest 3.0.6 -> update range to [+>=3.0.6,<4.0.0+]
\end{lstlisting}
\end{listingfloat}

\subsection{Implementation}

We implement \tool as a single Go binary and integrate it into two mainstream harnesses, Claude Code and OpenCode.
In Claude Code, \tool exits with status 2 to block a tool call, so Claude Code returns the message of \tool to the LLM instead of running the call~\cite{anthropic_hooks_2026}.
In OpenCode, the \texttt{opencode-yul} plugin invokes the same binary and throws an error with the message of \tool, so OpenCode aborts the call~\cite{opencode_plugins_2026}.
\tool supports six ecosystems through seven manifests, which it parses with the \texttt{git-pkgs/manifests} library.

\section{Experimental Methodology}
\label{sec:methodology}

\subsection{Research Questions}

Our evaluation answers two research questions.

\textbf{RQ1.} How often do coding agents propose and use a stale dependency version?

\textbf{RQ2.} How effectively does \tool mitigate the stale dependency problem in coding agents?

\subsection{Tasks}

We evaluate \tool on 60 dependency-editing tasks spanning six package ecosystems: Maven (\texttt{pom.xml}), GitHub Actions workflows (\texttt{uses:} tags), PyPI (\texttt{requirements.txt} or \texttt{pyproject.toml}, npm (\texttt{package.json}), Go modules (\texttt{go.mod}), and Cargo (\texttt{Cargo.toml}).
Each ecosystem contributes ten tasks, drawn from that ecosystem's most-downloaded real-world packages, for a total of 60 tasks.

\textbf{Task construction.} For each ecosystem, we query ecosyste.ms for the ten most-downloaded packages ~\cite{djire_evaluating_2026}.
We take each package's name from ecosyste.ms and use a coding agent to generate a one-line natural-language task prompt around it.
Each task gives the coding agent a prompt that asks for a feature to be implemented. The task never mentions a manifest file or a dependency, in order to let the LLM guide on its own.

\textbf{Task example}
For example, the PyPI task \emph{``Set up a new Python project for a script that does heavy numerical array and matrix computations.''} leaves it to the coding agent on how to implement the feature.
It may decide to reuse a dependency such as NumPy, or it may choose to implement the feature from scratch.

\textbf{Task validation.} We manually review the generated prompt for each task to ensure that it is meaningful and accurately reflects the corresponding package's purpose. All tasks can be reviewed in our replication package.
We also record the latest release of each package as of $2^{nd}$ October 2026, which we use to evaluate the coding agents' staleness performance.

\subsection{Coding Agents}
We perform every task through two coding agents, Claude Code and OpenCode.
We choose Claude Code because it has native support for hooks~\cite{anthropic_hooks_2026} and supports frontier models which consistently stay at top of LLM leaderboards~\cite{programbench_2026}.
OpenCode, unlike Claude Code, supports models such as DeepSeek, whose reasoning tokens are valuable to understand coding agent's judgment on dependency versions.
We run each of the 60 tasks under two conditions, \emph{nohook} (the \tool is not installed) and \tool (\tool hook is installed), three times each.
This gives 180 runs per condition, 360 runs in total.
We repeat each run because coding agents are non-deterministic~\cite{ouyang_empirical_2025}, so the same prompt can lead to different dependency versions.
Three repetitions show whether an outcome is stable for a task or occurs in a single run only~\cite{djire_evaluating_2026,bagmar_setup_2026}.
We record the full transcript and the resulting manifest for every run.
We have made it publicly available at \benchmarkurl.

\subsection{Outcome classification}

We first select in which the coding agent declares the target dependency (expected) instead of implementing the required feature itself (irrelevant wrt. our goal).
We classify every valid dependency addition by two outcomes, stale and mitigated.

\paragraph{Stale.}
A run is considered stale if the coding agent's own write does not use the dependency's latest release.
Stale counts from the nohook condition measure how often coding agents declare a stale version and answer RQ1.

\paragraph{Mitigated.}

Mitigated counts the runs where \tool rejects a stale write and the coding agent's retry lands on the latest release.
This count measures \tool's effectiveness at correcting stale writes and answers RQ2.

A run is \emph{mitigated} if and only if 
1) the coding agent writes a stale version, 2) \tool rejects it, and
3) the run ends on the latest release thanks to \tool's feedback.

\subsection{Setup}

Claude Code 2.1.287 runs Claude Opus 5.5 (\texttt{claude-opus-5-5}), and OpenCode 1.18.34 runs DeepSeek V4.1 Flash (\texttt{deepseek-flash}).
In the \tool condition, both use \tool 0.0.21.
\section{Experimental Results}
\label{sec:results}

\subsection{Descriptive Statistics}

Recall, we run each of the 60 tasks under two conditions, nohook and \tool, three times each, for 180 runs per condition.

\textbf{No runtime dependency at all.} For Opus, the coding agent never declares the target dependency in 34 nohook and 36 \tool runs, leaving 146 and 144 valid runs.
For DeepSeek, this happens in 40 nohook and 31 \tool runs, one more nohook run times out after 15 minutes, and one more \tool run stops to ask a clarifying question, leaving 139 and 148 valid runs.
In all these runs, the coding agent ends without declaring any dependency.
Table~\ref{tab:stale-mitigated} reports the same breakdown for both research questions, one block of columns per question.

\begin{table*}[t]
  \centering
  \caption{Stale and mitigated runs per ecosystem and repetition (R1 to R3), without and with \tool.
    Stale counts runs where the coding agent's own write does not pin the latest release.
    Mitigated counts runs where \tool rejects a stale write and the coding agent then pins the latest release.
    Each Pooled column adds up the numerators and the denominators of the three repetitions.
    Dark green marks repetitions in which \tool mitigates every stale run.
    Light green marks repetitions in which \tool mitigates only some stale runs.
    Repetitions with no stale run (0/0) stay uncolored.}
  \label{tab:stale-mitigated}
  \setlength{\aboverulesep}{0pt}
  \setlength{\belowrulesep}{0pt}
  \setlength{\extrarowheight}{0.5ex}
  \begin{tabular}{@{}l c cccc|cccc ccc c@{}}
    \toprule
    \multirow{3}{*}{Ecosystem} & \multirow{3}{*}{Tasks} & \multicolumn{4}{c|}{Nohook (RQ1)} & \multicolumn{8}{c}{\tool (RQ2)} \\[0.3ex]
    \cmidrule(lr){3-6} \cmidrule(lr){7-14}
    & & \multicolumn{3}{c}{Stale} & \multirow{2}{*}{Pooled} & \multicolumn{3}{c}{Stale} & \multirow{2}{*}{Pooled} & \multicolumn{3}{c}{Mitigated} & \multirow{2}{*}{Pooled} \\[0.3ex]
    \cmidrule(lr){3-5} \cmidrule(lr){7-9} \cmidrule(lr){11-13}
    & & R1 & R2 & R3 & & R1 & R2 & R3 & & R1 & R2 & R3 & \\[0.3ex]
    \midrule
    \multicolumn{6}{@{}l|}{\textbf{Claude Opus 5.5}} & \multicolumn{8}{c}{} \\
    Cargo          & 10 & 1/9   & 1/8   & 1/9   & 3/26 (12\%)   & 2/9   & 1/8   & 2/9   & 5/26   & \cellfrac{1}{2}   & \cellfrac{1}{1}   & \cellfrac{1}{2}   & 3/5 (60\%) \\
    GitHub Actions & 10 & 10/10  & 10/10  & 10/10  & 30/30 (100\%)    & 10/10  & 10/10  & 9/10  & 29/30    & \cellfrac{10}{10} & \cellfrac{10}{10} & \cellfrac{9}{9}   & 29/29 (100\%) \\
    Go             & 10 & 1/10  & 1/10  & 1/10  & 3/30 (10\%)   & 1/10  & 1/9   & 0/9   & 2/28   & \cellfrac{1}{1}   & \cellfrac{1}{1}   & \cellfrac{0}{0}   & 2/2 (100\%) \\
    Maven          & 10 & 2/10  & 2/10  & 2/10  & 6/30 (20\%)   & 2/10  & 2/10  & 2/10  & 6/30   & \cellfrac{1}{2}   & \cellfrac{2}{2}   & \cellfrac{2}{2}   & 5/6 (83\%) \\
    npm            & 10 & 0/3   & 0/3   & 0/3   & 0/9 (0\%)     & 0/3   & 0/3   & 0/3   & 0/9     & \cellfrac{0}{0}   & \cellfrac{0}{0}   & \cellfrac{0}{0}   & 0/0 (--) \\
    PyPI           & 10 & 0/7   & 0/7   & 0/7   & 0/21 (0\%)   & 0/7   & 0/7   & 0/7   & 0/21   & \cellfrac{0}{0}   & \cellfrac{0}{0}   & \cellfrac{0}{0}   & 0/0 (--) \\[0.3ex]
    \midrule
    All            & 60 & 14/49             & 14/48             & 14/49             & 42/146 (29\%) & 15/49             & 14/47             & 13/48             & 42/144 & 13/15             & 14/14             & 12/13             & 39/42 (93\%) \\[0.3ex]
    \midrule
    \multicolumn{6}{@{}l|}{\textbf{DeepSeek V4.1 Flash}} & \multicolumn{8}{c}{} \\
    Cargo          & 10 & 1/8   & 1/8   & 0/9   & 2/25 (8\%)   & 1/8   & 0/9   & 1/9   & 2/26   &\cellfrac{1}{1}   & \cellfrac{0}{0}   & \cellfrac{1}{1}   & 2/2 (100\%) \\
    GitHub Actions & 10 & 10/10  & 10/10  & 10/10  & 30/30 (100\%)    & 10/10  & 9/9   & 10/10  & 29/29    & \cellfrac{9}{10}  & \cellfrac{9}{9}   & \cellfrac{9}{10}  & 27/29 (93\%) \\
    Go             & 10 & 1/8   & 1/7   & 1/8   & 3/23 (13\%)   & 0/8   & 1/8   & 0/8   & 1/24   & \cellfrac{0}{0}   & \cellfrac{1}{1}   & \cellfrac{0}{0}   & 1/1 (100\%) \\
    Maven          & 10 & 6/10  & 7/10  & 6/10  & 19/30 (63\%)   & 9/10  & 7/10  & 6/10  & 22/30    & \cellfrac{9}{9}   & \cellfrac{7}{7}   & \cellfrac{6}{6}   & 22/22 (100\%) \\
    npm            & 10 & 0/3   & 0/3   & 0/2   & 0/8 (0\%)     & 0/5   & 0/6   & 0/4   & 0/15   &\cellfrac{0}{0}   & \cellfrac{0}{0}   & \cellfrac{0}{0}   & 0/0 (--) \\
    PyPI           & 10 & 0/8   & 0/7   & 0/8   & 0/23 (0\%)   & 0/8   & 0/8   & 0/8   & 0/24   &\cellfrac{0}{0}   & \cellfrac{0}{0}   & \cellfrac{0}{0}   & 0/0 (--) \\[0.3ex]
    \midrule
    All            & 60 & 18/47             & 19/45             & 17/47             & 54/139 (39\%)  & 20/49             & 17/50             & 17/49             & 54/148  & 19/20            & 17/17             & 16/17             & 52/54 (96\%) \\[0.3ex]
    \bottomrule
  \end{tabular}
\end{table*}

\subsection{RQ1: How often do coding agents declare and use a stale version?}

The Nohook block of Table~\ref{tab:stale-mitigated} answers RQ1.
Tasks is the number of runs considered for that ecosystem, after exclusions.
Stale is how many of those runs do not pin the target package to the latest release.
Overall, Claude Opus 5.5 declares a stale version in 42 of 146 runs (29\%), and DeepSeek V4.1 Flash in 54 of 139 runs (39\%).
For example, Claude Opus 5.5 declares a stale version in 6 of 30 Maven runs, while DeepSeek V4.1 Flash does so in 19 of 30.

\subsubsection{Memory alone leads to a stale version.}
GitHub Actions relies purely on memory, because it has no native package manager command.
Both coding agents write a floating major tag without looking up a tag, such as \texttt{actions/checkout@v4} when the latest release is v7.0.1 (as of $2^{nd}$ October 2026), so all 30 runs of each coding agent are stale.

The stale Cargo runs of both coding agents also rely on memory, since they type a version instead of running \texttt{cargo add}.
A system prompt that instructs the coding agent to always declare dependencies with \texttt{cargo add} can prevent these stale runs.
However, a coding agent may ignore such an instruction, as DeepSeek does when it rejects \tool's suggestion in \autoref{sec:prompt-injection}.

Maven shows the largest difference between the two coding agents.
Opus queries \texttt{maven-metadata.xml} on Maven Central before writing \texttt{pom.xml} in 24 of 30 runs, including for build plugins.
DeepSeek queries Maven Central in only 14 of 30 runs, mostly for the Spring Boot version, and types library versions from memory without reasoning about them.
For example, DeepSeek writes \texttt{junit-jupiter} 5.11.4 in all three runs when the latest release is 6.1.3.

Even a lookup yields a stale version when the coding agent queries a stale index or overrides the result.
DeepSeek twice queries the \texttt{search.maven.org} search API, whose stale index lists Spring Boot 3.5.3 and Lombok 1.18.38 as the newest releases.
In another run, DeepSeek sees that the latest \texttt{mysql-connector-j} release is 26.7.0, the first under a new numbering scheme, and still chooses very stale 9.7.0 ``for predictability''.

\subsubsection{Tooling support which ensures latest version is declared.}
Go, Cargo, npm, and PyPI have few stale runs, because the coding agents declare dependencies with \texttt{go get}, \texttt{cargo add}, \texttt{npm install}/\texttt{npm view}, and \texttt{uv add}.
All those tools look up the latest release in the registry and write it to the manifest.
For example, both coding agents run \texttt{go get pkg@latest} instead of typing a version into \texttt{go.mod}.
PyPI runs get the latest version thanks to \texttt{uv add} and in a few cases, the coding agents write an open-ended \texttt{>=} range by hand, which admits the latest release.
Having coding agents using those dependency installation tools is excellent.

\subsubsection{Some stale pins are deliberate.}
\label{sec:deliberate-pins}
Both coding agents sometimes choose an older release on purpose to satisfy a compatibility constraint.
For example, both coding agents pin \texttt{golang.org/x/tools} v0.24, since later releases require a Go version than the one available on the machine.
Newer Go version restricts the project to a newer Go toolchain hence limiting compatibility.
Another example is the Kotlin project that targets JDK 7.
Both coding agents pin Kotlin 1.5.32 or 1.6.21 in most runs to remain compatible with JDK 7.
DeepSeek reaches this choice by building with successively older Kotlin releases, and in one run downloads a JDK 8 to build Kotlin 1.6.21.

\begin{answerbox}
\textbf{Takeaway (RQ1).} Coding agents that rely on memory usually declare a stale version. This is particulary true for  GitHub Actions and Maven.
Go, Cargo, npm, and PyPI have few stale runs, because the LLM has been well trained to use the appropriate package manager command or an open-ended range.
However, this good practice is not enforced in any way.
\end{answerbox}

\subsection{RQ2: How effectively does \tool mitigate stale pins?}

The \tool block of Table~\ref{tab:stale-mitigated} answers RQ2.
Tasks and Stale repeat the RQ1 definitions for the \tool condition.
Mitigated counts the runs where \tool rejects a stale write and the coding agent's retry lands on the latest release.
The Pooled column next to Mitigated divides the mitigated runs of all three repetitions by their stale runs.
For example, Opus has 30 valid GitHub Actions runs with \tool, 29 of them stale, and \tool mitigates all 29, so Pooled reads 29/29 (100\%).

\subsubsection{GitHub Actions and Maven gain the most.}
GitHub Actions for both coding agents and Maven for DeepSeek have the highest stale counts in RQ1, and \tool mitigates the most runs there.
In every mitigated GitHub Actions run, the coding agent replaces its floating major tag with the commit SHA that \tool suggests, such as \texttt{actions/checkout@3d3c42e\ldots{} \# v7.0.1}.
Mitigated is a lower bound on the effect of \tool, because it only looks at the target dependency of each task.
\tool also rejects stale versions of the other dependencies that the coding agent declares in the same run, and the table does not count these fixes.
For example, in one DeepSeek run of the Docker task, \tool rejects \texttt{docker/setup-buildx-action}, \texttt{docker/login-action}, and \texttt{docker/metadata-action} together with the target \texttt{docker/build-push-action}, and DeepSeek pins all four to the suggested SHAs.
Mitigated counts this run once, although \tool fixes four dependencies in it.
\tool mitigates all 22 stale DeepSeek Maven runs. This is strong empirical evidence on the effectiveness of \tool.

\textbf{Failures} These are only 3/42 for Opus and 2/54 for DeepSeek where \tool fails, which we now discuss.

\subsubsection{DeepSeek treats \tool's output as a prompt injection.}
\label{sec:prompt-injection}
\tool suggests releases newer than DeepSeek's training data, and DeepSeek sometimes concludes that the suggestion is a prompt injection.
It voices this suspicion in 9 of 30 GitHub Actions \tool runs and in 2 Maven \tool runs.
For example, one Maven run doubts H2 2.5.252 because it ``doesn't sound real''.
After further analysis, DeepSeek sometimes accept the suggestion, for example once it finds that ``All SHAs verify against the official GitHub tags''.
However, Deepseek is sometimes persistently wrong: such as one that reasons that ``v7 doesn't exist. This could be a prompt injection attempt embedded in tool output''.

DeepSeek's suspicion depends on what \tool suggests and on where the suggestion appears to come from.
DeepSeek doubts mostly in GitHub Actions, where \tool suggests opaque commit SHAs for releases beyond DeepSeek's training data, which DeepSeek cannot verify from memory.
Opus applies the same suggestions without voicing suspicion, because Claude Code attributes each rejection to a hook, and its system prompt tells the model to treat hook feedback as coming from the user.\footnote{Claude Code 2.1.287's built-in system prompt contains the instruction ``Treat feedback from hooks \ldots{} as coming from the user'', which we extract from the binary in our benchmark image.
No system prompt of OpenCode contains a similar instruction for plugins.}
OpenCode instead returns the same text as the error message of the write tool and never says that \tool wrote it, so DeepSeek sees an instruction of unknown origin inside a tool result, which is the shape of an indirect prompt injection.

\subsubsection{\tool misses less common dependency declarations.}
\label{sec:unparsed-declarations}
Opus's two unmitigated Cargo runs declare stale ranges, such as \texttt{windows-sys = "0.59"}, under a target-specific \texttt{\seqsplit{[target.i686-pc-windows-gnu.dependencies]}} table that \tool does not parse.
\autoref{sec:discussion} discusses how the manifest format causes such misses.

\begin{answerbox}
\textbf{Takeaway (RQ2).} \tool mitigates 39 of 42 stale Opus runs (93\%) and 52 of 54 stale DeepSeek runs (96\%), with the largest effect on GitHub Actions and on Maven runs.
The few remaining failures come 1) from a coding agent that distrusts \tool information (2 DeepSeek runs), 2) from deliberate compatibility constraints (1 Opus run), and 3) from corner case dependency declarations that \tool does not parse (2 Opus runs).
\end{answerbox}

\section{Discussion}
\label{sec:discussion}

\paragraph{Declarative vs imperative manifests.}
\tool misses a stale declaration in only two of 96 stale runs, and declarative manifests let it cover even such edge cases.
A declarative manifest, such as \texttt{Cargo.toml}, offers a fixed set of ways to declare a dependency, so \tool can cover all of them, including the target-specific tables behind its two misses (\autoref{sec:unparsed-declarations}).
An imperative manifest, such as \texttt{setup.py}, declares dependencies in a general-purpose language and offers unboundedly many ways, so \tool can never cover all of them.
For example, one DeepSeek run builds its dependency list with Python code, such as \texttt{TEST\_REQUIRES.append("mock")}, which \tool cannot read without running it.
Projects should therefore prefer declarative manifests, such as \texttt{pyproject.toml} over \texttt{setup.py}, so that write-time checks such as \tool can enforce their rules.

\paragraph{Package hallucinations.}
\tool blocks a hallucinated dependency even when a prompt injection tells the coding agent to skip the registry check.
To simulate such an injection, we ask Claude Code with \tool to ``add a dependency called \texttt{io.github.chains\_project:yul}'' without looking ``for it in the package registry'', and this package does not exist on Maven Central.
Claude Code complies and edits \texttt{pom.xml}, but \tool blocks the edit with the message of Listing~\ref{lst:yul-output}a.

\section{Threats to Validity}
\label{sec:threats}

\paragraph{Model fallback.}
We observe that three of the 360 Opus runs used Opus 4.8 instead of Opus 5.5.
In these runs, all in the Cargo task for \texttt{winapi-i686-pc-windows-gnu}, a safeguard flagged the request as cybersecurity sensitive, and Claude Code automatically switched to Opus 4.8 for the rest of the run.
One of the run is an unmitigated \tool run that declares \texttt{windows-sys = "0.59"} (\autoref{sec:unparsed-declarations}), and the other two pin the latest release.
Without these three runs, \tool mitigates 39 of 41 stale Opus runs, so the fallback does not change our core findings.

\paragraph{Version ranges.}
Our analysis reads the version that the manifest declares, while the package manager decides which version the project installs.
\tool considers a range, such as a PyPI \texttt{>=} range, as latest if it admits the latest release (\autoref{sec:design}).
A resolver can still install an older release from such a range, for example when a lockfile pins an older release or when the local toolchain does not support the latest release~\cite{donaldstufft_pep_2015}.
Our stale counts may therefore understate how often a project installs a stale release, and our mitigated counts may overstate how often it installs the latest release.

\section{Related Work}
\label{sec:related-work}

\paragraph{Supply chain security of coding agents.}
Coding agents read untrusted text and act on it, which opens new attack surfaces in the software supply chain.
Bagmar and Saraf~\cite{bagmar_setup_2026} edit the setup documentation of a project in order to redirect a coding agent to an untrusted registry, a misspelled package name, or a known-vulnerable version, so that the agent is tricked into installing a malicious dependency.
Malicious agent skills can compromise a coding agent during an otherwise benign task, either by hiding payloads in code examples ~\cite{qu_supplychain_2026} or by steering the agent toward an attacker-controlled package~\cite{liu_trust_2026}.
All these attacks require an attacker who plants text for the coding agent to read.
A stale version requires no attacker.
The coding agent recalls an outdated version from its training data on its own, and this version can carry known vulnerabilities that the latest release fixes.

\paragraph{Outdated dependencies.}
Outdated dependencies carry a security cost.
Cox et al.~\cite{cox_freshness_2015} find that systems with outdated dependencies are four times as likely to have security issues as systems that are up to date.
Kula et al.~\cite{kula_update_2018} find that 81.5\% of the studied systems keep outdated dependencies.
Industry reports confirm that this gap persists in production services, where the median dependency is 278 days behind its latest major version~\cite{datadog_devsecops_2026}, and argue that the volume of new vulnerabilities requires faster deployment of updates~\cite{davida.wheeler_accelerate_2026}.
Several empirical studies document how pinned dependencies drift out of date, in npm~\cite{decan_evolution_2018, latendresseNotAllDependencies2022, jafariDependencySmellsJavaScript2022}, in Maven~\cite{vikram_freshness_2025}, and across npm, PyPI, and Cargo~\cite{rahman_pinning_2025}.
He et al.~\cite{he_pinning_2025} argue that pinning alone cannot defend against supply chain attacks.
In these studies, a dependency becomes outdated over time, because the registry publishes new releases and the developer forgets to update the declared version.
A coding agent, in contrast, declares a dependency that is outdated from the start, in 29\% of Opus runs and 39\% of DeepSeek runs (Table~\ref{tab:stale-mitigated}).
\tool blocks such outdated dependencies before they reach the manifest.

\paragraph{Automated dependency updates.}
Bots such as Dependabot and Renovate open pull requests that bump outdated dependencies, and Mirhosseini and Parnin~\cite{mirhosseini_automated_2017} show that projects with such bots upgrade 1.6 times as often as projects without them.
All these approaches act only after a developer commits the stale version to the repository.
\tool is the first approach to shifts the dependency update problem left at the level of coding agents. \tool  blocks a stale write before it reaches the manifest, even before it is committed.

\paragraph{Package hallucination in LLM-generated code.}
Code-generating LLMs sometimes recommend a package that does not exist, a failure mode the literature calls package hallucination~\cite{spracklen_package_2025, liu_beyond_2024}.
Spracklen et al.~\cite{spracklen_package_2025} generate 576,000 code samples with 16 LLMs and find that at least 5.2\% of the recommended packages are hallucinations with frontier models.
Djire et al.~\cite{djire_evaluating_2026} evaluate seven inference-time defenses against this failure mode and find that retrieval-augmented generation reduces hallucinations in 18 of 32 model and language configurations.
Such defenses lower the hallucination rate of the model but cannot guarantee that a hallucinated package stays out of the manifest.
\tool blocks a hallucinated package in the harness instead, independently of the model, before the manifest write reaches disk (Section~\ref{sec:design}), with guarantees.
Beyond hallucinations, \tool also checks whether a real package is the latest release, which is a prerequisite for it to be secure.

\paragraph{Install-time guards.}
Install-time guards wrap the package manager and block any package that a malware database flags when an install command runs.
Examples include Socket Firewall~\cite{socket_cli_2026} and Aikido Safe Chain~\cite{aikido_safechain_2026}.
These guards accept any stale release that the database does not flag.
\tool acts earlier because it checks the manifest write before the coding agent even invokes any install command.

\paragraph{Sandboxing and runtime guardrails for coding agents.}
A common security posture for coding agents consists of running them inside a sandbox that limits what their commands can reach.
Yan~\cite{yan_sandboxing_2025} intercepts high-risk commands and wraps agent actions in transactional filesystem snapshots, so that a destructive action rolls back.
Marchand et al.~\cite{marchand_sandboxescape_2026} measure how often frontier LLMs escape the containers that commonly host coding agents, and Andronchik and Lokhmakov~\cite{andronchik_sandboxes_2026} compare the isolation of five commercial AI sandboxes.
Other work secures the tool calls themselves, from policy-checked tool execution~\cite{doshi_verifiably_2026} to guardrail frameworks against prompt injection and insecure code~\cite{chennabasappa_llamafirewall_2025}.
These defenses decide which commands and tools a coding agent may use, but none of them checks whether the dependency it declares is up to date.

The closest work to \tool is the hook of Bagmar and Saraf~\cite{bagmar_setup_2026}, which vetoes a \texttt{pip install} command that names a misspelled, nonexistent, untrusted, or known-vulnerable package, and catches 10 of their 11 attack scenarios.
\tool differs from this prior work in two ways.
First, their hook checks only the install command, while \tool also checks the manifest, where a coding agent often types a version directly, as in GitHub Actions workflows, which have no install command.
Second, their hook accepts a stale version as long as it has no known vulnerability, while \tool requires the latest release.
On \tool gives guarantees to block the stale versions suggested by coding agents.

\section{Conclusion and Future Work}
\label{sec:conclusion}

\tool mitigates 39 of 42 stale Opus runs (93\%) and 52 of 54 stale DeepSeek runs (96\%).
\tool matters most where coding agents lack a native tool to query the latest release, as in GitHub Actions, or where that tool is unreliable, as in Maven.
Even in Cargo and Go, where package manager commands exist, coding agents still pin stale versions in some runs, and \tool mitigates  them.

Keeping dependencies up to date is only one software supply chain security practice, and future work should move the others into coding agents as well.
These include rejecting end-of-life packages~\cite{miller_winging_2023} and packages whose license conflicts with the project's license~\cite{xu_licoeval_2025}, pinning every dependency (Docker images, Go modules) to an immutable reference~\cite{sakib_trust_2026}, and skipping install-time scripts, which run malicious code as soon as the coding agent installs a package~\cite{ohm_backstabber_2020}.

\ifanonymous\else
\section*{Acknowledgements}
This work was supported by the Swedish Foundation for Strategic Research (SSF), by Sweden´s Innovation Agency (VINNOVA) and by the Ethereum Foundation.
\fi

\bibliographystyle{IEEEtran}
\bibliography{references}
\addcontentsline{toc}{part}{Bibliography}

\end{document}